\documentclass[conference]{IEEEtran}
\IEEEoverridecommandlockouts
\usepackage{cite}
\usepackage{amsmath,amssymb,amsfonts}
\usepackage{algorithmic}
\usepackage{graphicx}
\usepackage{textcomp}
\usepackage{xcolor}
\usepackage{balance}

\usepackage{subcaption}

\usepackage{pgfplots}
\pgfplotsset{compat=newest}
\pgfplotsset{plot coordinates/math parser=false}

\def\BibTeX{{\rm B\kern-.05em{\sc i\kern-.025em b}\kern-.08em
    T\kern-.1667em\lower.7ex\hbox{E}\kern-.125emX}}
\begin{document}

\title{Deterministic Equivalent of the Log-Euclidean Distance between Sample Covariance Matrices
\thanks{The work of
 Xavier Mestre was supported by the Spanish ministry of economic affairs and digital transformation and NextGeneration EU grant UNICO-5G I+D/AROMA3D-Earth (TSI-063000-2021-69) and by Grant 2021 SGR 00772 funded by the Generalitat de Catalunya. 
 The work of 
Roberto Pereira has been funded by the grant CHIST-ERA-20-SICT-004 (SONATA) by PCI2021-122043-2A/AEI/10.13039/501100011033. }
}

\author{\IEEEauthorblockN{Xavier Mestre, Roberto Pereira}
\IEEEauthorblockA{\textit{Centre Tecnològic de Telecomunicacions de Catalunya (CERCA-CTTC)} \\
Av. Carl Friedrich Gauss 7, 08860 Castelldefels, Spain 
}
 }

\maketitle

\begin{abstract}
Log-Euclidean distances are commonly used to quantify the similarity between positive definite matrices using geometric considerations. This paper analyzes the behavior of this distance when it is used to measure closeness between independent sample covariance matrices. A closed form expression is given for the deterministic equivalent of such distance, which asymptotically approximates the actual distance in the large observation regime (both sample size and observation dimension grow to infinity at the same rate). The deterministic equivalent can be used to analyze the performance of the log-Euclidean metric when compared to other commonly used metrics such as the Euclidean norm or the symmetrized Kullback-Leibler divergence. 
\end{abstract}


\section{Introduction}

Covariance matrices constitute a very powerful descriptor of multivariate data and are therefore employed in multiple applications, ranging from
clustering, 
classification, 
estimation
and many other statistical sciences. 
In a number of these contexts, one needs to  quantify the closeness of two covariance matrices by means of an appropriate distance. 
Recent approaches have focused on the study of distances that exploit the fact that covariance matrices naturally belong to the Riemmann manifold of positive definite matrices \cite{lhuang2017riemannian,shinohara2010covariance,li2013riemannian_euclidean,Barachant13,shi2019riemannian}. 
Initial works focused on the (squared) affine-invariant Riemannian metric which, given two covariance matrices $\mathbf{R}_1$, $\mathbf{R}_2$, is defined as 
\begin{equation} \label{eq:defAI}
d^{AI}_M = \frac{1}{M} \mathrm{tr}\left[\log^2(\mathbf{R}_2^{-1/2}\mathbf{R}_1\mathbf{R}_2^{-1/2})\right].
\end{equation} 
Note here that the logarithm is applied matrix-wise (i.e. to the eigenvalues).
Unfortunately, despite its appealing analytic considerations, this metric is computationally complex to implement.

More recent contributions have focused on developing alternative metrics that retain the desirable geometric properties of positive definite matrices while offering improved computational efficiency. This is the case of the (square) log-Euclidean distance proposed in \cite{arsigny06} in the context of diffusion tensor imaging, which is defined as
\begin{equation} 
\label{eq:defLE}
{d}_{M}^{LE} =\frac{1}{M}\mathrm{tr}\left[  \left(  \log\mathbf{R}_{1}%
-\log\mathbf{R}_{2}\right)  ^{2}\right]. 
\end{equation}
The log-Euclidean metric was originally derived by endowing the manifold of positive definite matrices with an appropriate Lie group structure, together with a logarithmic scalar multiplication that gives the essential properties of a vector space \cite{arsigny07}. 
Contrary to the affine-invariant metric, the log-Euclidean distance is more amenable from the computational complexity,  has a closed form solution for its (Fréchet) mean and always yields a positive definite Gaussian kernel~\cite{Jayasumana15}. Hence, it is often the distance of choice when comparing different covariance matrices \cite{ilea2018covariance,li2013log,Jayasumana15,wang2012covariance}. 

Now, an important challenge in practical approaches is the fact that covariance matrices are generally unknown. Consequently, the inherent distances must be estimated from the corresponding data. This becomes particularly challenging in situations where the number of available samples is not much larger than the corresponding observation dimension, for instance, in non-stationary environments. 
In these situations, it becomes crucial to ensure that the distance estimators are consistent (i.e., correctly approximate the distance between the true/population covariance matrices) when the number of samples per observation dimension is finite. This is typically the regime considered in random matrix theory, which has recently been quite useful to provide consistent estimators of these distances when both the sample size and the observation dimension increase without bound at the same rate. 
More specifically, these tools have been used in \cite{couillet2019random} to propose a consistent estimator of the affine-invariant metric in (\ref{eq:defAI}) and more recently in \cite{pereira_icassp23} to derive a consistent estimator of the log-Euclidean metric in (\ref{eq:defLE}). 

An important drawback of the above consistent estimators is the fact that they are only defined in the oversampled regime, that is when the number of samples is larger than the observation dimension. Unfortunately, in fast changing scenarios, one often only has access to a limited number of samples and the above estimators are no longer applicable.  Another problem is the computational complexity associated with their implementation, since it typically involves solving multiple polynomial equations in addition to the conventional eigendecomposition operations. To overcome these difficulties one may consider a more naive approach consisting in simply replacing the covariance matrix by their sample estimates, i.e. using the plug-in estimators. 
Differently from the consistent estimators above, these traditional plug-in estimators do not necessarily converge to the true distance and only approximate these values. Nonetheless, we emphasize that in many fast-changing scenarios it is often the only available option.  For the case of the log-Euclidean metric one would consider the plug-in estimator
\begin{equation} \label{eq:defLEsampled}
    \hat{d}_{M}^{LE} =\frac{1}{M}\mathrm{tr}\left[  \left(  \log\hat{\mathbf{R}}_{1}
-\log\hat{\mathbf{R}}_{2}\right)  ^{2}\right] 
\end{equation}
where $\hat{\mathbf{R}}_{j}$, $j \in \{1,2\}$ are the sample covariance matrices (SCM) obtained from the observations. The above metric can trivially be extended to the case where the SCMs are singular (undersampled regime), simply by considering the logarithm of the positive eigenvalues only and leaving the zero eigenvalues intact. Admittedly, this type of generalization is not supported by Riemmann geometry considerations but nonetheless seems reasonable enough from the algebraic perspective and is relatively easy to implement. For this reason, this will be the generalization to the undersampled regime that will be considered in this paper. The main objective of the paper is to derive a deterministic quantity that asymptotically approximates (\ref{eq:defLEsampled}) when both the sample size and the observation dimension are large but comparable in magnitude.
This deterministic equivalent can be obtained from the true covariance matrices and may provide insights on the  behavior or the log-Euclidean distance in real-world applications.

\section{Statistical model and deterministic equivalent}
We consider two sets of $M$-dimensional observations of size $N_{1}$ and $N_{2}$ respectively. If we denote by $\mathbf{Y}_{j}$, $j \in \{1,2\}$ the $M\times N_{j}$ matrix containing the observations of the $j$th set and if we assume that these observations have zero mean, one can express the SCM as $\mathbf{\hat{R}}_{j}={N_{j}}^{-1}\mathbf{Y}_{j}\mathbf{Y}_{j}^{H}$. 
We will make the following assumptions:

\noindent \textbf{(As1)} For $j\in\{1,2\}$ the matrix of observations
$\mathbf{Y}_{j}$ can be expressed as 
$
\mathbf{Y}_{j}=\mathbf{R}_{j}^{1/2}\mathbf{X}_{j}
$
where $\mathbf{X}_{j}$ is an $M \times N_j$ matrix of independent and identically distributed entries with zero mean and unit variance. 

\noindent \textbf{(As2)} The different eigenvalues of $\mathbf{R}_{j}$ are denoted $0<\gamma_{1}^{(j)}<\ldots<\gamma_{\bar{M}_{j}}^{(j)}$ ($j \in \{1,2\}$)\ and have multiplicity $K_{1}^{(j)},\ldots,K_{\bar{M}_{j}}^{(j)}$, where $\bar{M}_{j}$ is the total number of distinct eigenvalues. All these quantities may vary with $M$ \ but we always have $\inf_{M}\gamma_{1}^{(j)}>0$ and $\sup_{M}
\gamma_{\bar{M}_j}^{(j)}<\infty$.

\noindent \textbf{(As3)} The quantities $N_{1}$ and $N_{2}$ depend on $M$, that is
$N_{1}=N_{1}(M)$ and $N_{2}=N_{2}(M)$. Furthermore, when $M\rightarrow\infty$ we have, for $j\in\{1,2\}$, $N_{j}(M)\rightarrow\infty$ in a way that $N \neq M$ and $M/N_{j}
\rightarrow c_{j}$ for some constant $0<c_{j}<\infty$ such that $c_{j}\neq1$.

In order to analyze the behavior of $\hat{d}_{M}^{LE}$ under the above assumptions,
we need some definitions that are useful in the context of random matrix theory. Consider the function of complex variable $\omega_{j}\left(  z\right)$, given by one of the solutions to the
polynomial equation
\begin{equation}
z=\omega_{j}\left(  z\right)  \left(  1-\frac{1}{N_{j}}\sum_{m=1}^{\bar{M}_{j}}K_{m}^{(j)}\frac{\gamma_{m}^{(j)}}{\gamma_{m}^{(j)}-\omega_{j}\left(
z\right)  }\right)  . \label{eq:defw(z)}
\end{equation}
More specifically, if $z\in\mathbb{C}^{+}$ (upper complex semiplane),
$\omega_{j}\left(  z\right)  $ is the only solution of the above equation located in $\mathbb{C}
^{+}$. If $z\in\mathbb{C}^{-}$(lower complex semiplane), $\omega_{j}\left(
z\right)  $ is the only solution in $\mathbb{C}^{-}$. Finally, if $z$ is real
valued, $\omega_{j}\left(  z\right)  $ is defined as the only real valued
solution such that
\begin{equation}
\frac{1}{N_{j}}\sum_{m=1}^{\bar{M}_{j}}K_{m}^{(j)}\left(  \frac{\gamma_{m}^{(j)}}{\gamma_{m}^{(j)}-\omega_{j}\left(  z\right)  }\right)  ^{2}<1.
\label{eq:conditionwrealvalued}
\end{equation}

Let us consider the resolvent $\mathbf{Q}_{j}(\omega)  =\left(  \mathbf{R}_{j}-\omega\mathbf{I}_{M}\right)  ^{-1}$, which is well defined for $\omega \in \mathbb{C}$ outside the set of eigenvalues of $\mathbf{R}_j$. Based on this, we define $\bar{\mathbf{Q}}_{j}(z) = \omega(z)/z \mathbf{Q}_{j}(\omega(z))$ 
 for $j\in\{1,2\}$. Now, it can be shown \cite[Corollary 1]{pereira23tsp} that under $\textbf{(As1)}-\textbf{(As3)}$ we have $\hat{d}_M^{LE}-\bar{d}_M^{LE} \rightarrow 0$ with probability one, where $\bar{d}_M^{LE}$ is usually referred to as the deterministic equivalent of the original random distance $\hat{d}_M^{LE}$. This deterministic equivalent is defined as 
\begin{align*}
\bar{d}_{M}^{LE}  & =\frac{-1}{ 4\pi^2 }\oint
\nolimits_{\mathcal{N}_{1}}\oint\nolimits_{\mathcal{Z}_{2}}\log^{2}%
(z_{1})\frac{1}{M}\mathrm{tr}\left[  \mathbf{\bar{Q}}_{1}(z_{1})\mathbf{\bar
{Q}}_{2}(z_{2})\right]  dz_{1}dz_{2}\\
& \frac{1}{2\pi^2}\oint\nolimits_{\mathcal{N}%
_{1}}\oint\nolimits_{\mathcal{N}_{2}}\log(z_{1})\log(z_{2})\frac{1}%
{M}\mathrm{tr}\left[  \mathbf{\bar{Q}}_{1}(z_{1})\mathbf{\bar{Q}}_{2}%
(z_{2})\right]  dz_{1}dz_{2}\\
& +\frac{-1}{4\pi^2}\oint\nolimits_{\mathcal{Z}%
_{1}}\oint\nolimits_{\mathcal{N}_{2}}\log^{2}(z_{2})\frac{1}{M}\mathrm{tr}%
\left[  \mathbf{\bar{Q}}_{1}(z_{1})\mathbf{\bar{Q}}_{2}(z_{2})\right]
dz_{1}dz_{2}%
\end{align*}
where $\mathcal{N}_{j}$ and $\mathcal{Z}_{j}$ are both
negatively oriented contours enclosing the interval $[\theta_{j}^{-}%
-\epsilon,\theta_{j}^{+}+\epsilon]$ for some small $\epsilon>0$ such that
$\epsilon<\theta_{j}^{-}$, where $\theta_{j}^{-}=\inf_{M}(1-\sqrt{M/N_{j}%
})^{2}\gamma_{1}^{(j)}$ and $\theta_{j}^{+}=\sup_{M}(1-\sqrt{M/N_{j}}%
)^{2}\gamma_{\bar{M}_{j}}^{(j)}$ and where $\mathcal{Z}_{j}$ encloses zero and
$\mathcal{N}_{j}$ does not. The main objective of this paper is to derive a
closed form analytical expression for $\bar{d}_{M}^{LE}$ by essentially solving the above integrals. 

\section{A closed form expression for $\bar{d}_M^{LE}$}

Let us now come up with a closed form expression for $\bar{d}_M^{LE}$. We begin by noticing that
\[
\frac{1}{2\pi\mathrm{j}}\oint\nolimits_{\mathcal{Z}_{j}}\mathbf{\bar{Q}}%
_{j}(z_{j})dz_{j}=\mathbf{I}_{M}.
\]
This can be proven by applying a change of variable $z \mapsto \omega=\omega(z)$ and using conventional Cauchy integration (see \cite{pereira23tsp} for details). The above identity allows us to express 
\begin{equation} \label{eq:asymptoticEquivalent}
\bar{d}_{M}^{LE}=\alpha^{(1)}-2\frac{1}{M}\mathrm{tr}\left[  \boldsymbol{\Theta}^{(1)}%
\boldsymbol{\Theta}^{(2)}\right]  +\alpha^{(2)}%
\end{equation}
where we have defined
\begin{align}
\alpha^{(j)}  & =\frac{1}{2\pi\mathrm{j}}\oint\nolimits_{\mathcal{N}_{j}}%
\log^{2}(z_{j})\frac{1}{M}\mathrm{tr}\left[  \mathbf{\bar{Q}}_{j}%
(z_{j})\right]  dz_{j}  \label{eq:integralAlpha}\\
\boldsymbol{\Theta}^{(j)}  & =\frac{1}{2\pi\mathrm{j}}\oint\nolimits_{\mathcal{N}_{j}}%
\log(z_{j})\mathbf{\bar{Q}}_{j}(z_{j})dz_{j}. \label{eq:integralTheta}
\end{align}
Hence, in order to obtain a closed form expression for the deterministic
equivalent $\bar{d}_{M}$ it is sufficient to evaluate these two integrals.

To present the result, we need some additional notation. First of all, let us
consider $\mu_{0}^{(j)}<\mu_{1}^{(j)}<\ldots<\mu_{\bar{M}_j}^{(j)}$ the $\bar{M}_j +1$ solutions to the equation 
\begin{equation} \label{eq:defmuseq}
\mu\left(  1-\frac{1}{N_{j}}\sum_{r=1}^{\bar{M}_{j}}K_{r}^{(j)}\frac
{\gamma_{r}^{(j)}}{\gamma_{r}^{(j)}-\mu}\right)  =0.
\end{equation}
One can  see that $\mu_{0}^{(j)}<0=\mu_{1}^{(j)}$ in the undersampled regime and $\mu
_{0}^{(j)}=0<\mu_{1}^{(j)}$ in the oversampled regime. Let us also define, for  $m=1,\ldots,\bar{M}_j$  the quantities
\begin{align}
 \gamma_{m,0}^{(j)} & = \gamma_{m}^{(j)}-\mu_{0}^{(j)}\\
 \mu_{m,0}^{(j)} & = \mu_{m}^{(j)}-\mu_{0}^{(j)} 
\end{align}
and observe that $\gamma_{m,0}^{(j)}=\gamma_{m}^{(j)}$, $\mu_{m,0}^{(j)}=\mu_{m}^{(j)}$ 
in the oversampled regime ($N_j>M$). Finally, define $\Gamma_{0}^{(j)}=\Gamma_{j} (\mu_{0}^{(j)} )$ where
\[
\Gamma_{j}\left(  \omega\right)  =\frac{1}{N_{j}}\sum_{r=1}^{\bar{M}_{j}}%
K_{r}^{(j)}\left(  \frac{\gamma_{r}^{(j)}}{\gamma_{r}^{(j)}-\omega}\right)
^{2}.
\]
Having introduced the necessary notation, we are now in the position of presenting a closed form evaluation of the two integrals in (\ref{eq:integralAlpha})-(\ref{eq:integralTheta}).  

More specifically, the matrix integral in (\ref{eq:integralTheta}) can be shown to take the form (see the Appendix for details)
\begin{equation} \label{eq:defTheta}
\boldsymbol{\Theta}^{(j)}=\sum_{k=1}^{\bar{M}_{j}}\beta_{k}^{(j)}\boldsymbol{\Pi}_{k}^{(j)}
\end{equation}
where $\boldsymbol{\Pi}_{k}^{(j)}$ is the orthogonal projection matrix onto the subspace spanned by the eigenvector(s) associated to the $k$th eigenvalue of $\mathbf{R}_j$ and where the coefficients $\beta_{k}^{(j)}$ take the form
\begin{multline*}
\beta_{k}^{(j)}   =\frac{\gamma_{k}^{(j)}}{\gamma_{k,0}^{(j)}}\left[  \log\left(  \gamma_{k,0}^{(j)}\right) +\log\left(
1-\Gamma_{0}^{(j)}\right)  -1\right] \\ +\sum_{\substack{m=1\\m\neq k}}^{\bar{M}_{j}}\frac{\gamma_{k}^{(j)}}
{\gamma_{k}^{(j)}-\gamma_{m}^{(j)}}\log  \frac{\gamma_{m,0}^{(j)}}{\gamma_{k,0}^{(j)}} 
 -\sum_{m=1}^{\bar{M}_{j}}\frac{\gamma_{k}^{(j)}}{\gamma_{k}^{(j)}-\mu_{m}^{(j)}}\log \frac
{\mu_{m,0}^{(j)}}{\gamma_{k,0}^{(j)}} .
\end{multline*}
The evaluation of the coefficient $\alpha^{(j)}$ is a bit more involved: details are omitted here due to space constraints but can be found in \cite{Mestre24logEuclidean}. It can be shown that the
integral can also be expressed in closed form as 
\begin{align*}
\alpha^{(j)} &  =2\frac{\min(N_j,M)}{M} + \frac{1}{M}\sum_{m=1}^{\bar{M}_j}K^{(j)}_{m}\left(\log^{2}  \gamma^{(j)}_{m,0} -2 \log\gamma^{(j)}_{m,0} \right) 
\\
&  + \frac{2}{M}\sum_{m=1}^{\bar{M}_j}\sum_{\substack{k=1\\k\neq m}}^{\bar{M}_j
}K^{(j)}_{m}\log\frac{\gamma^{(j)}_{m,0}}{\gamma^{(j)}_{k,0}}\log\frac{\gamma^{(j)}
_{m,0}}{\left\vert \gamma^{(j)}_{k}-\gamma^{(j)}_{m}\right\vert }\\
&-\frac{2}{M}
\sum_{m=1}^{\bar{M}_j}\sum_{k=1}^{\bar{M}_j}K^{(j)}_{m}\log\frac{\gamma^{(j)}_{m,0}}
{\mu^{(j)}_{k,0}}\log\frac{\gamma^{(j)}_{m,0}}{\left\vert \mu^{(j)}_{k}-\gamma^{(j)}_{m}\right\vert }\\
&  +\frac{2}{M}\sum_{m=1}^{\bar{M}_j}\sum_{k=1}^{\bar{M}_j
}K^{(j)}_{k}\left[  \Phi_{2}\left(  \frac{\gamma^{(j)}_{m,0}}{\gamma^{(j)}_{k,0}
}\right)  -\Phi_{2}\left(  \frac{\mu^{(j)}_{m,0}}{\gamma^{(j)}_{k,0}}\right)
\right]  \\
&+\alpha^{(j)}_{\text{os}}\delta_{M<N_j}+\alpha^{(j)}_{\text{us}}\delta_{M>N_j}
\end{align*}
where $\alpha^{(j)}_{\text{os}}$ (resp. $\alpha^{(j)}_{\text{us}}$) is a quantity that is present only in the oversampled (resp. undersampled) regime, and 
where we have introduced the function $\Phi_{2}(x)$, which is closely related to the di-logarithm. This function can
can be expressed as 
\begin{equation}
\Phi_{2}(x)=
\left\{
\begin{array}
[c]{ccc}%
\mathrm{Li}_{2}\left(  x\right)   &  & x<1\\
\frac{\pi^{2}}{3}-\frac{1}{2}\log^{2}x-\mathrm{Li}_{2}\left(  x^{-1}\right)
&  & x\geq1
\end{array}
\right.  \label{eq:definitionPhi(x)Li2}%
\end{equation}
and where $\mathrm{Li}_{2}\left(  x\right)  $ is the conventional
di-logarithm, that is
\[
\mathrm{Li}_{2}(x) = -\int_{0}^{x}\frac{\log\left\vert 1-y\right\vert }{y}dy.
\]
To complete the description of $\alpha^{(j)}$, we only need 
to specify the two quantities $\alpha^{(j)}_{\text{os}}$ and $\alpha^{(j)}_{\text{us}}$. The first can be simply expressed as 
\begin{align*}
\alpha^{(j)}_{\text{os}}  & =-\left(  \frac{N_j}{M}-1\right) \left[ \log^{2}\left(
1-\frac{M}{N_j}\right) - 2 \log\left(
1-\frac{M}{N_j}\right) \right] 
\\
& +\left(  \frac{N_j}{M}-1\right)   \sum_{m=1}^{\bar{M}_j}  \left[ \log^{2}
\gamma^{(j)}_{m}-\log^{2}\mu^{(j)}_{m}\right]
\end{align*}
whereas the second takes the slightly more involved form 
\begin{align*}
\alpha^{(j)}_{\text{us}}  & =\left(  1-\frac{N_j}{M}\right) \Bigg[ 2 \log\left\vert \mu^{(j)}_{0}\right\vert 
-\log^{2}\left\vert \mu^{(j)}
_{0}\right\vert \\ &- 2  \log\left\vert \mu^{(j)}
_{0}\right\vert \log\left(  \frac{M}{N_j}-1\right)  \\
& -2  \sum_{m=1}^{\bar{M}_j}\left[  \Phi_{2}\left(
\frac{\gamma^{(j)}_{m,0}}{\left\vert \mu^{(j)}_{0}\right\vert }\right)  -\Phi
_{2}\left(  \frac{\mu^{(j)}_{m,0}}{\left\vert \mu^{(j)}_{0}\right\vert }\right)
\right]  \\
& +2   \sum_{m=1}^{\bar{M}_j}\log
\gamma^{(j)}_{m,0}  \log\frac{\left\vert \mu^{(j)}_{0}\right\vert }
{\gamma_{m}^{(j)}}-2\sum_{m=2}^{\bar{M}_j}\log  \mu^{(j)}_{m,0}  \log
\frac{\left\vert \mu^{(j)}_{0}\right\vert }{ \mu^{(j)}_{m} } \Bigg].
\end{align*}
The above expressions provide a full description of the two quantities in (\ref{eq:integralAlpha})-(\ref{eq:integralTheta}) that can readily be used in (\ref{eq:asymptoticEquivalent}) to evaluate the asymptotic equivalent of the log-Euclidean metric in closed form. This allows to have a first order comparison of the behavior of this metric with other metrics, such as the Euclidean distance or the symmetrized Kullback-Leibler divergence (see \cite{pereira23tsp} for a deterministic equivalent of these two distances). 

\section{Numerical Evaluation}

\begin{figure}[tb]
\begin{subfigure}{0.9\textwidth}
    {\definecolor{mycolor1}{RGB}{44,160,44}
\definecolor{mycolor2}{rgb}{0.85000,0.32500,0.09800}%
\definecolor{mycolor3}{RGB}{31,119,180}

\definecolor{mycolor4}{rgb}{0.49400,0.18400,0.55600}%
\definecolor{mycolor5}{RGB}{31,119,180}%

\definecolor{darkgray176}{RGB}{176,176,176}
\definecolor{darkorange25512714}{RGB}{255,127,14}
\definecolor{forestgreen4416044}{RGB}{44,160,44}
\definecolor{lightgray204}{RGB}{204,204,204}
\definecolor{magenta}{RGB}{255,0,255}
\definecolor{orange2551868}{RGB}{255,186,8}
\definecolor{saddlebrown164660}{RGB}{164,66,0}
\definecolor{steelblue31119180}{RGB}{31,119,180}
\definecolor{saddlebrown164660}{RGB}{164,66,0}

\begin{tikzpicture} 
\begin{axis}[
hide axis,
width=0.8\linewidth,
height=0.3\linewidth,
at={(-1,0)},
xmin=10,
xmax=50,
ymin=0,
ymax=0.4,
legend style={fill opacity=1, draw opacity=1, text opacity=1, draw=lightgray204, legend columns=2,
    anchor=north,
    at={(0., 0)},
    font=\small,
    legend image post style={scale=0.6}
}
]

\addlegendimage{mycolor1, line width=2pt};
\addlegendentry{$N_1/M = 0.1, N_2/M=0.4$ \hspace{1em}};

\addlegendimage{mycolor2, line width=2pt};
\addlegendentry{$N_1/M = 0.8, N_2/M=0.8$ \hspace{1em}};

\addlegendimage{mycolor3, line width=2pt};
\addlegendentry{$N_1/M = 1.5, N_2/M=3$ \hspace{1em}};

\addlegendimage{mycolor4, line width=2pt};
\addlegendentry{$N_1/M = 2, N_2/M=8$ \hspace{1em}};

\coordinate (legend) at (axis description cs:-0.0,0.03);

\end{axis}

\end{tikzpicture}\vspace{1em}}
    \end{subfigure}
     \centering
     \hspace*{-0.15in} 
     \begin{subfigure}[b]{0.23\textwidth}
         \centering
%
%

\definecolor{mycolor1}{RGB}{44,160,44}
\definecolor{mycolor2}{rgb}{0.85000,0.32500,0.09800}%
\definecolor{mycolor3}{RGB}{31,119,180}

\definecolor{mycolor4}{rgb}{0.49400,0.18400,0.55600}%
\definecolor{mycolor5}{RGB}{31,119,180}%


%
\begin{tikzpicture}

\begin{axis}[%
width=0.75\linewidth,
height=1.0in,
at={(0in,1.9in)},
scale only axis,
xmin=2,
xmax=64,
xlabel style={font=\color{white!15!black}},
xmin=5,
xmax=60,
ymin=3,
ymax=7,
yminorticks=true,
axis background/.style={fill=white},
ylabel={Squared Distance},
xlabel style={text width=\textwidth, align=center},
xlabel={Growing $M$ }, 
align =center,
title={{Undersampled Regime \\  $\mathbf{R}_1 = \mathbf{R}_2$}}
]

\addplot [color=mycolor1, forget plot]
 plot [error bars/.cd, y dir=both, y explicit, error bar style={line width=0.5pt}, error mark options={line width=0.5pt, mark size=6.0pt, rotate=90}]
 table[row sep=crcr, y error plus index=2, y error minus index=3]{%
10	5.57504597421756	0.985203028428206	0.985203028428206\\
20	5.62203136895156	0.49666543469437	0.49666543469437\\
30	5.62598510617504	0.333838179478975	0.333838179478975\\
40	5.63543590698365	0.251443709842636	0.251443709842636\\
50	5.64027611517005	0.203669680583519	0.203669680583519\\
60	5.64616072193979	0.170509290191644	0.170509290191644\\
};

\addplot [color=mycolor1, dashdotted, forget plot]
  table[row sep=crcr]{%
10	5.6605610338051\\
20	5.6605610338051\\
30	5.6605610338051\\
40	5.6605610338051\\
50	5.6605610338051\\
60	5.6605610338051\\
};

\addplot [color=mycolor2, forget plot]
 plot [error bars/.cd, y dir=both, y explicit, error bar style={line width=0.5pt}, error mark options={line width=0.5pt, mark size=6.0pt, rotate=90}]
 table[row sep=crcr, y error plus index=2, y error minus index=3]{%
10	4.5162493880796	1.14065216752029	1.14065216752029\\
20	4.21535706717018	0.519300524424611	0.519300524424611\\
30	4.10846480759651	0.331170840918381	0.331170840918381\\
40	4.05983127662127	0.246267042004764	0.246267042004764\\
50	4.0257735396008	0.194265501636558	0.194265501636558\\
60	4.00217229930714	0.160176948684902	0.160176948684902\\
};
\addplot [color=mycolor2, dashdotted, forget plot]
  table[row sep=crcr]{%
10	3.89561470267826\\
20	3.89561470267827\\
30	3.89561470267827\\
40	3.89561470267827\\
50	3.89561470267826\\
60	3.89561470267826\\
};

\end{axis}


\begin{axis}[%
width=0.75\linewidth,
height=1.0in,
at={(0in,0in)},
scale only axis,
xmin=2,
xmax=64,
xlabel style={font=\color{white!15!black}},
ymin=0.5,
ymax=2.6,
yminorticks=true,
axis background/.style={fill=white},
legend style={at={(1.35in,0.7in)},anchor=north,
    legend columns=3, font=\small},
xlabel style={text width=\textwidth, align=center},
xlabel={Growing $M$ }, 
ylabel={Squared Distance},
align =center,
title={{Oversampled Regime \\  $\mathbf{R}_1 = \mathbf{R}_2$}}]

\addplot [color=mycolor3, forget plot]
 plot [error bars/.cd, y dir=both, y explicit, error bar style={line width=0.5pt}, error mark options={line width=0.5pt, mark size=6.0pt, rotate=90}]
 table[row sep=crcr, y error plus index=2, y error minus index=3]{%
10	1.70317865684115	0.45923604262236	0.45923604262236\\
20	1.56998721283775	0.212675191168801	0.212675191168801\\
30	1.52556315696883	0.136117274741136	0.136117274741136\\
40	1.50578159326882	0.100988969850299	0.100988969850299\\
50	1.49282745587399	0.0791667185491646	0.0791667185491646\\
60	1.48575292305168	0.0676368693400708	0.0676368693400708\\
};
\addplot [color=mycolor3, dashdotted, forget plot]
  table[row sep=crcr]{%
10	1.44312708035619\\
20	1.44312708035619\\
30	1.44312708035619\\
40	1.44312708035619\\
50	1.44312708035619\\
60	1.44312708035619\\
};

\addplot [color=mycolor4, forget plot]
 plot [error bars/.cd, y dir=both, y explicit, error bar style={line width=0.5pt}, error mark options={line width=0.5pt, mark size=6.0pt, rotate=90}]
 table[row sep=crcr, y error plus index=2, y error minus index=3]{%
10	0.93852353474514	0.238653230055272	0.238653230055272\\
20	0.877209562295512	0.112176779367207	0.112176779367207\\
30	0.854655310255562	0.0744885388188942	0.0744885388188942\\
40	0.843784577279349	0.0550824865871508	0.0550824865871508\\
50	0.837367160205527	0.043772671996933	0.043772671996933\\
60	0.833406973436072	0.0364823344558982	0.0364823344558982\\
};
\addplot [color=mycolor4, dashdotted, forget plot]
  table[row sep=crcr]{%
10	0.812708312594431\\
20	0.812708312594431\\
30	0.812708312594431\\
40	0.812708312594431\\
50	0.812708312594433\\
60	0.812708312594431\\
};

\end{axis}

\end{tikzpicture}%
     \end{subfigure}
     \hspace*{0.05in}
     \begin{subfigure}[b]{0.23\textwidth}
         \centering
%
%
\definecolor{mycolor1}{RGB}{44,160,44}
\definecolor{mycolor2}{rgb}{0.85000,0.32500,0.09800}%
\definecolor{mycolor3}{RGB}{31,119,180}

\definecolor{mycolor4}{rgb}{0.49400,0.18400,0.55600}%
\definecolor{mycolor5}{RGB}{31,119,180}%

\begin{tikzpicture}

\begin{axis}[%
width=0.75\linewidth,
height=1.0in,
at={(0in,1.9in)},
scale only axis,
xmin=2,
xmax=64,
xlabel style={font=\color{white!15!black}},
ymin=3,
ymax=7,
yminorticks=true,
axis background/.style={fill=white},
legend style={at={(1.35in,0.7in)},anchor=north,
    legend columns=3, font=\small},
xlabel style={text width=\textwidth, align=center},
xlabel={Growing $M$ }, 
align =center,
title={Undersampled Regime \\ $\mathbf{R}_1 \neq \mathbf{R}_2$},
]

\addplot [color=mycolor1, forget plot]
 plot [error bars/.cd, y dir=both, y explicit, error bar style={line width=0.5pt}, error mark options={line width=0.5pt, mark size=6.0pt, rotate=90}]
 table[row sep=crcr, y error plus index=2, y error minus index=3]{%
10	5.26864785363224	0.967871915092368	0.967871915092368\\
20	5.31604983879466	0.488289323761876	0.488289323761876\\
30	5.31837047078901	0.327767484117485	0.327767484117485\\
40	5.32597524373597	0.246692337891983	0.246692337891983\\
50	5.33113750277189	0.200080108920071	0.200080108920071\\
60	5.33695618755461	0.167419011046797	0.167419011046797\\
};
\addplot [color=mycolor1, dashdotted, forget plot]
  table[row sep=crcr]{%
10	5.3503752127958\\
20	5.3503752127958\\
30	5.3503752127958\\
40	5.3503752127958\\
50	5.3503752127958\\
60	5.3503752127958\\
};

\addplot [color=mycolor2, forget plot]
 plot [error bars/.cd, y dir=both, y explicit, error bar style={line width=0.5pt}, error mark options={line width=0.5pt, mark size=6.0pt, rotate=90}]
 table[row sep=crcr, y error plus index=2, y error minus index=3]{%
10	4.57035932466158	1.18131386276334	1.18131386276334\\
20	4.25818209115029	0.535505968160725	0.535505968160725\\
30	4.14588628567859	0.343320523005933	0.343320523005933\\
40	4.09612230483699	0.254545953591325	0.254545953591325\\
50	4.05961478950522	0.200071093966616	0.200071093966616\\
60	4.03529535503114	0.165761353532634	0.165761353532634\\
};
\addplot [color=mycolor2, dashdotted, forget plot]
  table[row sep=crcr]{%
10	3.92481138266841\\
20	3.92481138266841\\
30	3.92481138266841\\
40	3.92481138266841\\
50	3.9248113826684\\
60	3.9248113826684\\
};

\end{axis}


\begin{axis}[%
width=0.75\linewidth,
height=1.0in,
at={(0in,0in)},
scale only axis,
xmin=2,
xmax=64,
xlabel style={font=\color{white!15!black}},
xlabel style={text width=\linewidth, align=center},
xlabel={Growing $M$}, 
ymin=0.5,
ymax=2.5,
yminorticks=true,
axis background/.style={fill=white},
align =center,
title= {Oversampled Regime \\ $\mathbf{R}_1 \neq \mathbf{R}_2$},
]

\addplot [color=mycolor3, forget plot]
 plot [error bars/.cd, y dir=both, y explicit, error bar style={line width=0.5pt}, error mark options={line width=0.5pt, mark size=6.0pt, rotate=90}]
 table[row sep=crcr, y error plus index=2, y error minus index=3]{%
10	1.93760876359962	0.46269932306976	0.46269932306976\\
20	1.8154862463096	0.215256543834663	0.215256543834663\\
30	1.77327378360478	0.138274772256496	0.138274772256496\\
40	1.75497037155192	0.103288303747384	0.103288303747384\\
50	1.74370783797163	0.0808017432031783	0.0808017432031783\\
60	1.73634270980793	0.068885067628911	0.068885067628911\\
};
\addplot [color=mycolor3, dashdotted, forget plot]
  table[row sep=crcr]{%
10	1.6975589049325\\
20	1.6975589049325\\
30	1.6975589049325\\
40	1.6975589049325\\
50	1.6975589049325\\
60	1.6975589049325\\
};

\addplot [color=mycolor4, forget plot]
 plot [error bars/.cd, y dir=both, y explicit, error bar style={line width=0.5pt}, error mark options={line width=0.5pt, mark size=6.0pt, rotate=90}]
 table[row sep=crcr, y error plus index=2, y error minus index=3]{%
10	1.18312346630716	0.252987577232825	0.252987577232825\\
20	1.1296209711092	0.120182434544738	0.120182434544738\\
30	1.11048012413629	0.0799202188152704	0.0799202188152704\\
40	1.10026916239973	0.0588536235396564	0.0588536235396564\\
50	1.09524698170477	0.0472998106103182	0.0472998106103182\\
60	1.09155206172152	0.0396334563806181	0.0396334563806181\\
};
\addplot [color=mycolor4, dashdotted, forget plot]
  table[row sep=crcr]{%
10	1.07332667748523\\
20	1.07332667748523\\
30	1.07332667748523\\
40	1.07332667748523\\
50	1.07332667748523\\
60	1.07332667748523\\
};

\end{axis}

\end{tikzpicture}%
     \end{subfigure}
      \vspace{-1.5\baselineskip}
      \caption{Convergence of the log-Euclidean distance towards the asymptotic equivalent for different relative system dimensions. Solid lines represent simulated distances while dash-dotted lines show asymptotic equivalents. Vertical bars indicate the standard deviation of the simulated distance ($10^4$ realizations).}
    \label{fig:sims} 
    \vspace{-1\baselineskip}
\end{figure}
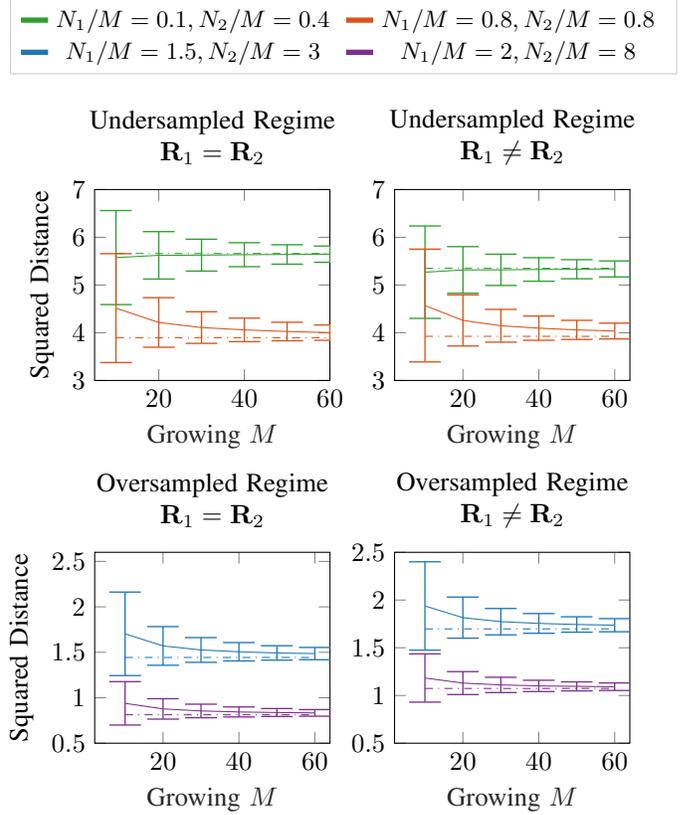

In order to illustrate the accuracy of the deterministic equivalents in a specific setting, we consider here a scenario where the actual (true) covariance $\mathbf{R}_1$ consists of four different eigenvalues $\{1, 6, 15, 25\}$  with relative multiplicities $\{0.1, 0.2, 0.3, 0.4\}$ respectively, whereas the covariance $\mathbf{R}_2$ has the same eigenvalues $\{ 1, 6, 15, 25\}$ with relative multiplicity $\{ 0.2, 0.2, 0.2, 0.4 \}$ respectively. We consider a system with large dimensions in which the observation dimension $M$ varies from $M=10$ to $M=60$ while keeping the ratios $M/N_1$ and $M/N_2$ fixed. The
SCMs
are built from Gaussian, circularly symmetric complex data with covariance equal to $\mathbf{R}_1$ and $\mathbf{R}_2$ respectively. We also consider the case where both sample covariance matrix are generated from $\mathbf{R}_1$, which we will indicate by $\mathbf{R}_1 = \mathbf{R}_2$. For each simulated $M$, the eigenvectors of the true covariance matrices are taken as the columns of random orthogonal matrices uniformly distributed on the corresponding Grassmann manifold. Figure~\ref{fig:sims} shows the convergence of the empirical  distance (obtained from multiple realizations of SCMs) towards the asymptotic equivalent (obtained from $\mathbf{R}_1, \mathbf{R}_2$) for different values of the ratios $N_1/M$ and $N_2/M$ in both the undersampled and the oversampled regimes. 
Results show that the deterministic equivalent (dashed lines)  provides a very accurate approximation of the observed random log-Euclidean distance  (solid lines), even for relatively low system dimensions. 
One can therefore rely on the deterministic equivalent as first order approximations of the behavior of the distance in a practical scenario.

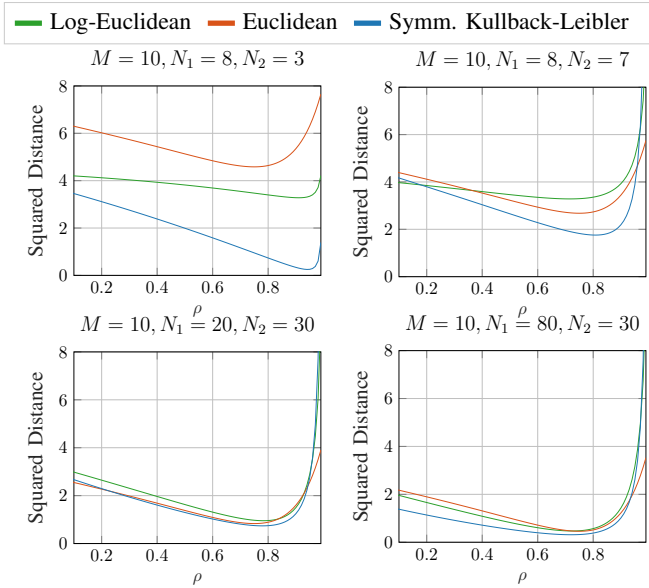
\begin{figure}[hbt]
\begin{subfigure}{0.9\textwidth}
     {
     \definecolor{mycolor1}{RGB}{44,160,44}
\definecolor{mycolor2}{rgb}{0.85000,0.32500,0.09800}%
\definecolor{mycolor3}{RGB}{31,119,180}

\definecolor{mycolor4}{rgb}{0.49400,0.18400,0.55600}%
\definecolor{mycolor5}{RGB}{31,119,180}%

\definecolor{darkgray176}{RGB}{176,176,176}
\definecolor{darkorange25512714}{RGB}{255,127,14}
\definecolor{forestgreen4416044}{RGB}{44,160,44}
\definecolor{lightgray204}{RGB}{204,204,204}
\definecolor{magenta}{RGB}{255,0,255}
\definecolor{orange2551868}{RGB}{255,186,8}
\definecolor{saddlebrown164660}{RGB}{164,66,0}
\definecolor{steelblue31119180}{RGB}{31,119,180}
\definecolor{saddlebrown164660}{RGB}{164,66,0}

\begin{tikzpicture} 
\begin{axis}[
hide axis,
width=0.8\linewidth,
height=0.3\linewidth,
at={(-1,0)},
xmin=10,
xmax=50,
ymin=0,
ymax=0.4,
legend style={fill opacity=1, draw opacity=1, text opacity=1, draw=lightgray204, legend columns=3,
    anchor=north,
    at={(0., 0)},
    font=\small,
    legend image post style={scale=0.6}
}
]

\addlegendimage{mycolor1, line width=2pt};
\addlegendentry{Log-Euclidean \hspace{1em}};

\addlegendimage{mycolor2, line width=2pt};
\addlegendentry{Euclidean \hspace{1em}};

\addlegendimage{mycolor3, line width=2pt};
\addlegendentry{Symm. Kullback-Leibler \hspace{1em}};

\coordinate (legend) at (axis description cs:0.97,0.03);

\end{axis}

\end{tikzpicture}
     }
     \end{subfigure}
     \centering
     \hspace*{-0.2in} 
     \begin{subfigure}[b]{0.55\textwidth}
         \centering
        \include{figs_tk/studyRho}
     \end{subfigure}
  \vspace{-4\baselineskip}
\caption{Comparison of the deterministic equivalents of 
the square log-Euclidean distance, the Euclidean distance and the symmetrized Kullback-Leibler divergence.}
    \label{fig:sims2} 
    \vspace{-1\baselineskip}
\end{figure}

To illustrate how the asymptotic equivalents can be used to assess the quality of a distance, we consider next another scenario in which the two covariance matrices $\mathbf{R}_1$ and $\mathbf{R}_2$ are built as Toeplitz matrix with the first row equal to $\left[\rho^0_j, \rho^1_j,\ldots,\rho^{M-1}_j\right]$, $j \in \{1,2\}$. More specifically, we fix $\rho_1 = 0.75$ and allow $\rho_2$ to take values between $0$ and $1$. Figure~\ref{fig:sims2} compares the deterministic equivalent of the log-Euclidean distance derived above with the asymptotic equivalents of two other metrics: the Euclidean distance and the symmetrized Kullback-Leibler divergence (see \cite{pereira23tsp} for an expression of these two quantities). Observe that in general the minimum of these deterministic equivalents does not need to coincide with the point for which the $\mathbf{R}_1 = \mathbf{R}_2$. 
Naturally, this is an undesired behavior that is likely to negatively affect any solution deployed in these scenarios. Such behaviors can be directly predicted from the first order approximations discussed throughout this work.

\section{Conclusions}
A closed form expression has been provided for the deterministic equivalent of the log-Euclidean distance between two independent sample covariance matrices. The approximation holds asymptotically when the number of samples of each sample covariance matrix grows to infinity at the same rate as the observation dimension. The result can be used to determine the behavior of this distance in practical finite-dimensional scenarios and to compare its performance against other metrics. 

\section*{Appendix \\ derivation of the expression for $\boldsymbol{\Theta}^{(j)}$}

In order to simplify the notation, we drop the dependence on the covariance
matrix index $j$ in this appendix. In order to deal with both the undersampled
and the oversampled regime in a natural way, we use the following trick. We
first notice that substracting the two sides of equations (\ref{eq:defw(z)}) and
(\ref{eq:defmuseq}) particularized to $\mu=\mu_{0}$ we obtain
\[
z=\left(  \omega(z)-\mu_{0}\right)  \left(  1-\Phi\left(  \omega(z)\right)
\right)
\]
where we have defined
\[
\Phi(\omega)=\frac{1}{N}\sum_{r=1}^{\bar{M}}K_{r}\frac{\gamma_{r}^{2}}{\left(
\gamma_{r}-\mu_{0}\right)  \left(  \gamma_{r}-\omega\right)  }.
\]
It can readily be seen that $\operatorname{Re}\left(  \omega(z)-\mu
_{0}\right)  >0$ for all $z$ inside $\mathcal{N}$, which shows that
$\log\left(  \omega(z)-\mu_{0}\right)  $ is analytic inside this contour.
Therefore by applying the conventional change of variable $z\mapsto\omega(z)$
we immediately see that (\ref{eq:defTheta}) can be transformed into
\begin{align}
\boldsymbol{\Theta} &  =\frac{1}{2\pi\mathrm{j}}\oint\nolimits_{\mathcal{N}_{\omega}}%
\frac{\omega\log\left(  \omega-\mu_{0}\right)  z^{\prime}(\omega)}{\left(
\omega-\mu_{0}\right)  \left(  1-\Phi\left(  \omega\right)  \right)
}\mathbf{Q}(\omega)d\omega\label{eq:Theta2Integr}\\
&  +\frac{1}{2\pi\mathrm{j}}\oint\nolimits_{\mathcal{N}_{\omega}}\frac
{\omega\log\left(  1-\Phi(\omega)\right)  z^{\prime}(\omega)}{\left(
\omega-\mu_{0}\right)  \left(  1-\Phi\left(  \omega\right)  \right)
}\mathbf{Q}(\omega)d\omega\nonumber
\end{align}
where $\mathcal{N}_{\omega}=\omega\left(  \mathcal{N}\right)  $, $z^{\prime
}(\omega)=1-\Gamma(\omega)$ and where we use the conventional short hand
notation $z=z(\omega)$ to denote the mapping in (\ref{eq:defw(z)}). We can
easily compute the second integral since it has all the singularities inside
$\mathcal{N}_{\omega}$ except for the pole at $\mu_{0}$ (see \cite[Fig. 1]{pereira23tsp}). One can therefore enlarge by adding the corresponding residue at
\ $\mu_{0}$ (which turns out to be equal to $\mu_{0}\log\left(  1-\Gamma
_{0}\right)  \mathbf{Q}(\mu_{0})$), so that one can write
\begin{multline*}
\frac{1}{2\pi\mathrm{j}}\oint\nolimits_{\mathcal{N}_{\omega}}\frac{\omega
\log\left(  1-\Phi(\omega)\right)  z^{\prime}(\omega)}{\left(  \omega-\mu
_{0}\right)  \left(  1-\Phi\left(  \omega\right)  \right)  }\mathbf{Q}%
(\omega)d\omega \\=\mu_{0}\log\left(  1-\Gamma_{0}\right)  \mathbf{Q}(\mu_{0})\\
+\frac{1}{2\pi\mathrm{j}}\oint\nolimits_{\mathcal{\bar{N}}_{\omega}}%
\frac{\omega\log\left(  1-\Phi(\omega)\right)  z^{\prime}(\omega)}{\left(
\omega-\mu_{0}\right)  \left(  1-\Phi\left(  \omega\right)  \right)
}\mathbf{Q}(\omega)d\omega
\end{multline*}
where $\mathcal{\bar{N}}_{\omega}$ now encloses both $\mathcal{N}_{\omega}$
and $\mu_{0}$. As a result of this enlargement, it turns out that all the
singularities of the integrand of the second term are inside the contour. One
can therefore apply a second change of variables $\omega\mapsto\zeta\left(
\omega\right)  =\omega^{-1}$ to show that
\begin{multline*}
\frac{1}{2\pi\mathrm{j}}\oint\nolimits_{\mathcal{\bar{N}}_{\omega}}
\frac{\omega\log\left(  1-\Phi(\omega)\right)  z^{\prime}(\omega)}{\left(
\omega-\mu_{0}\right)  \left(  1-\Phi\left(  \omega\right)  \right)
}\mathbf{Q}(\omega)d\omega \\ =\frac{1}{2\pi\mathrm{j}}\oint\nolimits_{\mathcal{Z}
_{0}}\log\left(  1-\Phi(\xi^{-1})\right)  \times\\
\times\frac{1-\frac{1}{N}\sum_{r=1}^{\bar{M}}K_{r}\left(  \frac{\xi\gamma_{r}
}{\xi\gamma_{r}-1}\right)  ^{2}}{1-\frac{1}{N}\sum_{r=1}^{\bar{M}}K_{r}
\frac{\xi\gamma_{r}}{\xi\gamma_{r}-1}}\left(  \xi\mathbf{R-I}_{M}\right)
^{-1}\frac{1}{\xi}d\xi=0
\end{multline*}
where $\mathcal{Z}_{0}$ is a negatively oriented contour that only encloses
zero. This means that we can disregard the second term in
(\ref{eq:Theta2Integr}) and simply focus on the evaluation of the first
integral. This one can easily solve it by evaluating the residues at the poles
$\gamma_{r},\mu_{r}$, $r=1,\ldots,\bar{M}$, that is
\begin{align*}
\boldsymbol{\Theta} &  =\log\left(  1-\Gamma_{0}\right)  \mu_{0}\mathbf{Q}(\mu_{0})\\
&  +\sum_{m=1}^{\bar{M}}\sum_{\substack{k=1\\k\neq m}}^{\bar{M}}\frac
{\gamma_{m}\log \gamma_{m,0} }{\gamma_{k}-\gamma_{m}}
\Pi_{k}-\sum_{m=1}^{\bar{M}}\sum_{k=1}^{\bar{M}}\frac{\mu_{m}\log
\mu_{m,0}  }{\gamma_{k}-\mu_{m}}\Pi_{k}\\
&  +\sum_{m=1}^{\bar{M}}\frac{\log \gamma_{m,0}
}{K_{m,0}}\left(  \sum_{r=1}^{\bar{M}}\frac{K_{r,0}\gamma_{r}}{\gamma_{r}
-\mu_{m}}-\sum_{\substack{r=1\\r\neq m}}^{\bar{M}}\frac{K_{r,0}\gamma_{r}
}{\gamma_{r}-\gamma_{m}}\right)  \Pi_{m}\\
&  +\sum_{m=1}^{\bar{M}}\frac{\mu_{0}}{  \gamma_{m,0}}
\log \gamma_{m,0}  \Pi_{m}-\sum_{m=1}^{\bar{M}}
\frac{\gamma_{m}}{ \gamma_{m,0}  }\Pi_{m}
\end{align*}
where $K_{m,0}=K_{m}\gamma_{m}/\left(  \gamma_{m}-\mu_{0}\right)  $, and where
we have used the fact that $\Phi(\mu_{m})=1$ for all $\mu_{m}$ inside the
contour (namely $\mu_{1},\ldots,\mu_{\bar{M}}$). We can simplify the above
expression using the identities (they can be proven following the
same steps as in \cite[Lemma 1]{Mestre08tsp}):
\begin{align*}
\sum_{\substack{r=1\\r\neq m}}^{\bar{M}}\frac{K_{r,0}\gamma_{r}}{\gamma
_{r}-\gamma_{m}}-\sum_{r=1}^{\bar{M}}\frac{K_{r,0}\gamma_{r}}{\gamma_{r}%
-\mu_{m}}  & =\sum_{\substack{k=1\\k\neq m}}^{\bar{M}}\frac{K_{m,0}\gamma_{m}%
}{\gamma_{m}-\gamma_{k}}-\sum_{r=1}^{\bar{M}}\frac{K_{m,0}\gamma_{m}}%
{\gamma_{m}-\mu_{r}}\\
\prod_{m=1}^{\bar{M}}\frac{\mu_{m,0}}{\gamma_{m,0}}  &
= 1-\Gamma_{0}.
\end{align*}
A direct application of these two identities directly leads to the expression of $\boldsymbol{\Theta}$ presented above.

\balance

\bibliographystyle{IEEEtran}
\bibliography{./bib/IEEEbib}

\begin{thebibliography}{10}
\providecommand{\url}[1]{#1}
\csname url@samestyle\endcsname
\providecommand{\newblock}{\relax}
\providecommand{\bibinfo}[2]{#2}
\providecommand{\BIBentrySTDinterwordspacing}{\spaceskip=0pt\relax}
\providecommand{\BIBentryALTinterwordstretchfactor}{4}
\providecommand{\BIBentryALTinterwordspacing}{\spaceskip=\fontdimen2\font plus
\BIBentryALTinterwordstretchfactor\fontdimen3\font minus \fontdimen4\font\relax}
\providecommand{\BIBforeignlanguage}[2]{{%
\expandafter\ifx\csname l@#1\endcsname\relax
\typeout{** WARNING: IEEEtran.bst: No hyphenation pattern has been}%
\typeout{** loaded for the language `#1'. Using the pattern for}%
\typeout{** the default language instead.}%
\else
\language=\csname l@#1\endcsname
\fi
#2}}
\providecommand{\BIBdecl}{\relax}
\BIBdecl

\bibitem{lhuang2017riemannian}
Z.~Huang and L.~Van~Gool, ``A {Riemannian} network for {SPD} matrix learning,'' in \emph{Thirty-first AAAI conference on artificial intelligence}, 2017.

\bibitem{shinohara2010covariance}
Y.~Shinohara, T.~Masuko, and M.~Akamine, ``Covariance clustering on riemannian manifolds for acoustic model compression,'' in \emph{2010 IEEE International Conference on Acoustics, Speech and Signal Processing}.\hskip 1em plus 0.5em minus 0.4em\relax IEEE, 2010, pp. 4326--4329.

\bibitem{li2013riemannian_euclidean}
Y.~Li and K.~M. Wong, ``Riemannian distances for signal classification by power spectral density,'' \emph{IEEE Journal of Selected Topics in Signal Processing}, vol.~7, no.~4, pp. 655--669, 2013.

\bibitem{Barachant13}
A.~Barachant, S.~Bonnet, M.~Congedo, and C.~Jutten, ``Classification of covariance matrices using a {Riemannian}-based kernel for {BCI} applications,'' \emph{Neurocomputing}, vol. 112, pp. 172--178, 2013, advances in artificial neural networks, machine learning, and computational intelligence.

\bibitem{shi2019riemannian}
Y.~Shi, J.~Xu, and K.~Wong, ``{A Riemannian Distance Approach to MIMO Radar Signal Design},'' in \emph{ICASSP 2019-2019 IEEE International Conference on Acoustics, Speech and Signal Processing (ICASSP)}.\hskip 1em plus 0.5em minus 0.4em\relax IEEE, 2019, pp. 4170--4174.

\bibitem{arsigny06}
V.~Arsigny, P.~Fillard, X.~Pennec, and N.~Ayache, ``Log-{Euclidean} metrics for fast and simple calculus on diffusion tensors,'' \emph{Magnetic Resonance in Medicine}, vol.~56, no.~2, pp. 411--421, 2006.

\bibitem{arsigny07}
------, ``Geometric means in a novel vector space structure on symmetric positive-definite matrices,'' \emph{SIAM Journal on Matrix Analysis and Applications}, vol.~29, no.~1, pp. 328--347, 2007.

\bibitem{Jayasumana15}
S.~Jayasumana, R.~Hartley, M.~Salzmann, H.~Li, and M.~Harandi, ``Kernel methods on {Riemannian} manifolds with {Gaussian} {RBF} kernels,'' \emph{IEEE Transactions on Pattern Analysis and Machine Intelligence}, vol.~37, no.~12, pp. 2464--2477, 2015.

\bibitem{ilea2018covariance}
I.~Ilea, L.~Bombrun, S.~Said, and Y.~Berthoumieu, ``Covariance matrices encoding based on the log-euclidean and affine invariant riemannian metrics,'' in \emph{Proceedings of the IEEE Conference on Computer Vision and Pattern Recognition Workshops}, 2018, pp. 393--402.

\bibitem{li2013log}
P.~Li, Q.~Wang, W.~Zuo, and L.~Zhang, ``Log-{E}uclidean kernels for sparse representation and dictionary learning,'' in \emph{Proceedings of the IEEE international conference on computer vision}, 2013, pp. 1601--1608.

\bibitem{wang2012covariance}
R.~Wang, H.~Guo, L.~S. Davis, and Q.~Dai, ``Covariance discriminative learning: A natural and efficient approach to image set classification,'' in \emph{2012 IEEE conference on computer vision and pattern recognition}.\hskip 1em plus 0.5em minus 0.4em\relax IEEE, 2012, pp. 2496--2503.

\bibitem{couillet2019random}
R.~Couillet, M.~Tiomoko, S.~Zozor, and E.~Moisan, ``Random matrix-improved estimation of covariance matrix distances,'' \emph{Journal of Multivariate Analysis}, vol. 174, p. 104531, 2019.

\bibitem{pereira_icassp23}
R.~Pereira, X.~Mestre, and D.~Gregoratti, ``Consistent estimators of a new class of covariance matrix distances in the large dimensional regime,'' in \emph{ICASSP 2023 - 2023 IEEE International Conference on Acoustics, Speech and Signal Processing (ICASSP)}, 2023, pp. 1--5.

\bibitem{pereira23tsp}
\BIBentryALTinterwordspacing
------, ``Asymptotics of distances between sample covariance matrices,'' \emph{Accepted at IEEE Transactions on Signal Processing}, 2023. [Online]. Available: \url{https://ieeexplore.ieee.org/document/10443725}
\BIBentrySTDinterwordspacing

\bibitem{Mestre24logEuclidean}
\BIBentryALTinterwordspacing
X.~Mestre and R.~Pereira, ``Asymptotics of the log-{Euclidean} distance between sample covariance matrices,'' 2024. [Online]. Available: \url{https://cloud.cttc.es/index.php/s/q8Yys7GXzbNm5rR}
\BIBentrySTDinterwordspacing

\bibitem{Mestre08tsp}
X.~Mestre, ``On the asymptotic behavior of the sample estimates of eigenvalues and eigenvectors of covariance matrices,'' \emph{IEEE Transactions on Signal Processing}, vol.~56, no.~11, pp. 5353--5368, 2008.

\end{thebibliography}

\end{document}